\documentclass[11pt]{amsart}
\usepackage{fullpage}
\usepackage{amsthm,amsmath,amssymb}
\usepackage[foot]{amsaddr}

\usepackage{enumitem}
\usepackage{booktabs}
\usepackage{graphicx}
\usepackage{subcaption}
\usepackage{tikz}

\usepackage[colorlinks=true,linkcolor=red,breaklinks=true,citecolor=blue,urlcolor=cyan]{hyperref}

\graphicspath{{./figs/}} 

\newtheorem{theorem}{Theorem}[section]
\newtheorem{proposition}[theorem]{Proposition}

\theoremstyle{remark}

\title{Partial Identification of the Average Treatment Effect with Stochastic Counterfactuals and Discordant Twins}

\author{Brian Knaeble}
\address{Department of Computer Science, Utah Valley University, Orem, UT}
\email{bknaeble@uvu.edu}

\author{Braxton Osting}
\address{Department of Mathematics, University of Utah, Salt Lake City, UT}
\email{osting@math.utah.edu}

\author{Placede Tshiaba}
\address{Department of Mathematics, University of Utah, Salt Lake City, UT}
\email{placede.tshiaba@utah.edu}


\begin{document}

\begin{abstract}
We develop a novel approach to partially identify causal estimands, such as the average treatment effect (ATE), from observational data.
To better satisfy the stable unit treatment value assumption (SUTVA) we utilize stochastic counterfactuals within a propensity-prognosis model of the data generating process. For more precise identification we utilize knowledge of discordant twin outcomes as evidence for randomness in the data generating process. 
Our approach culminates with a constrained optimization problem; the solution gives upper and lower bounds for the ATE. 
We demonstrate the applicability of our introduced methodology with three example applications.
\end{abstract}

\maketitle

\section{Introduction}
Is marijuana a gateway drug? Does smoking cause chronic obstructive pulmonary disease? Does diabetes cause strokes? Although these are important causal questions, they are difficult to answer conclusively from observational data since the exposures are not randomly assigned. 
Accurate causal interpretation of an observed association is required for successful intervention policy formation.

Rubin emphasized potential outcomes in an observational setting, defining individual causal effects \cite[Section 2.8]{Rubin1974,Imbens2015}. Rubin's causal model is generally applicable but often utilized within a framework of deterministic counterfactuals; c.f. \cite{VR}. This presents a problem because in the examples above the outcomes occur long after the exposures. For that reason the potential outcomes are likely better modeled stochastically rather than deterministically. We address this problem with a framework \cite{Knaeble2023} that  utilizes propensity probabilities and also 
counterfactual prognosis probabilities. 

There is still the problem of a lack of random exposure assignment. A hallmark of randomness is unpredictability. If identical twins are known to often be discordant for the exposure, then the process giving rise to exposure is likely more stochastic. Within our propensity-prognosis model of the data generating process, knowledge of discordant twins can be transported and utilized to partially identify the average treatment effect (ATE).

%
%
The contributions of this paper are described as follows. (i) We derive an inequality that relates a measure of concordance with the variance of propensity probabilities; see Proposition~\ref{probandwiseprop}. (ii) We formulate a constrained optimization problem for the purpose of partial identification of the ATE; see~\eqref{probs}. 
(iii) We describe a method of approximately solving the optimization problem with a linear program.

Our paper is organized as follows. 
In Section \ref{methods}, we give some background on our propensity-prognosis model of the data generating process and twin studies.
In Section \ref{results}, we present a mathematical formulation of the problem of partial identification. 
In Section \ref{applications}, we describe three example applications of our methodology. 
We conclude in Section~\ref{discussion} with a discussion. 
A proof of Proposition \ref{probandwiseprop}, found in Section \ref{results}, is given in Appendix \ref{AppA}.
\section{Background: a propensity-prognosis model and twins!}
\label{methods}
Here, we describe our propensity-prognosis model of the data generating process. 
We then introduce some concepts from twin studies  and describe an important assumption; see (\ref{assumption}).

\subsection{A propensity-prognosis model}
For individual $i$ we write $e_i$ to indicate exposure (or treatment) and $d_i$ to indicate a dichotomous outcome (or disease). For individual $i$ we write $\pi_i$ for propensity probability of exposure, $r_{0i}$ for prognosis probability of the outcome in the absence of exposure, and $r_{1i}$ for prognosis probability of the outcome in the presence of exposure. For individual $i$ causation is present if $r_{0i}\neq r_{1i}$. Here is a formal summary: 
\begin{align*}
e_i &\sim \textrm{Bernoulli}(\pi_i) \\
d_i(e_i=0) &\sim \textrm{Bernoulli}(r_{0i}), 
\quad  \textrm{and} \\ 
d_i(e_i=1) & \sim \textrm{Bernoulli}(r_{1i}).
\end{align*}
%

In the language of Imbens and Rubin \cite{Imbens2015} processes are assumed to be individualistic (except within twin pairs) and probabilistic, and we have assumed for each $i$ that $e_i$ does not depend on $(d_i(e_i=0),d_i(e_i=1))$ (see the latter requirement of SUTVA in \cite[Section 3.2]{Imbens2015}). We do not assume that the processes are unconfounded, i.e. we do not accept the assumption of strongly ignorable treatment assignment \cite[Section 3.4, Definition 3.6]{Imbens2015}. We measure $\{e_i,d_i\}_{i=1}^N$ while $\{\pi_i,r_{0i},r_{1i}\}_{i=1}^N$ are latent and unknown. For each $i$, define \emph{expected risk} to be \[r_i:=\pi_i r_{1i}+(1-\pi_i)r_{0i}.\] 
%

Denote the open unit cube by $C:=\{(\pi,r_0,r_1)\in \mathbb{R}^3\colon 0<\pi,r_0,r_1<1\}$ and the space of probability distributions on $C$ by $\mathcal P(C)$. 
We write $\mu\in\mathcal P(C)$ for the unknown distribution of real $(\pi,r_0,r_1)$ values. 
We define the \textit{average treatment effect} as
\begin{equation}
    \label{ATE}
\textrm{ATE}:=\int_C (r_1-r_0) d\mu.
\end{equation}
The ATE is a real but unknown causal estimand. Typically, the average treatment effect is defined as $\mathbb{E}(y_1-y_0)$ where only one of the potential outcomes $(y_0,y_1)$ is observed for each individual. In our definition, both of $(r_0,r_1)$ are latent and unobserved for each individual, but nevertheless we will still be able to bound our ATE as defined in (\ref{ATE}). For individual $i$ the difference $r_{1i}-r_{0i}$ is their change in risk caused by the exposure. The ATE is a meaningful causal estimand because knowledge of it can facilitate formation of intervention policy on a population of apparently homogeneous individuals. Other causal estimands are briefly discussed in Section \ref{discussion}.

\subsection{Twin studies: concordance and an important assumption}
Studies of monozygotic twins often report measures of concordance \cite{Hagenbeek2023}. The pairwise rate of concordance is the proportion of affected twin pairs in which both members are affected, and the probandwise rate of concordance is the proportion of affected individuals among the co-twins of previously defined index cases \cite{Allen1967}. Our results are stated here using probandwise concordance, but this can be obtained from pairwise concordance  using \eqref{e:PCBC} in Appendix \ref{AppA}. Denote probandwise concordance by $BC$, and utilize a subscript to indicate which trait. 

Let $j$ index a population of $n$ pairs of identical twins, and let $i\in\{1,2\}$ index the individuals within any given pair, and let $\psi_{ji}$ denote propensity probability of the trait for individual $i$ in the $j$th pair. Propensity probabilities are functions of individual genomic variables and the state of the environment at the time of conception. Because the twins are identical and the propensity probabilities are defined at conception, we have for each $j$ that $\psi_{j1}=\psi_{j2}$. In reality it may turn out that both twins of a pair acquire the trait (event $A_j$), or it may turn out that both twins of the same pair avoid the trait (event $B_j$). The sum of the probabilities for those two events ($P(A_j)+P(B_j)$) is expected to be greater than $\psi_j^2+(1-\psi_j)^2$, because of common causes that affect two individuals sharing the same environment. It is true that a ``polarizing event'' could cause contrarian behavior and discordant outcomes. For instance, after being teased for looking alike, twins could intentionally make lifestyle choices to be different from each other, e.g. one could smoke and the other wouldn't. However, we think that polarizing events like that are much less likely than regular common causes from a shared environment, like chemical exposures in the womb \cite{Buckley2022,Braun2016}. It is therefore reasonable to assume 
\begin{subequations}\label{assumption}
\begin{align}\label{oneone}
\frac{1}{n} \sum_{j=1}^n P(A_j)+P(B_j)  
&\geq \frac{1}{2n} \sum_{j=1}^n \sum_{i=1}^2 \psi_{ji}^2+(1-\psi_{ji})^2 \\ 
\label{onetwo}
&= \frac{1}{N} \sum_{i=1}^N \psi_{i}^2+(1-\psi_{i})^2.
\end{align}
\end{subequations}  
The equality of line (\ref{onetwo}) holds if we assume that the twin sub population is representative of the general population. There is some support for that assumption \cite[p. 852]{Lykken1982,Hagenbeek2023}, but it may not hold for all recent studies \cite{Vitthala2008}.

\section{A mathematical formulation for the partial identification of the ATE}
\label{results}
In Section \ref{methods} we defined the probandwise concordance $BC$ and propensity probability $\psi$ of a trait that could be either the exposure $e$ or the outcome $d$. 
\begin{proposition} \label{probandwiseprop}
If the assumption of (\ref{assumption}) holds then for large $N$ we have
\begin{equation}
\label{assumption2}
\frac{1}{N} \sum_{i=1}^N(\psi_i-\bar{\psi})^2 
\leq \bar{\psi}(BC-\bar{\psi}),
\end{equation}
where $\bar{\psi} = \frac{1}{N}\sum_{i=1}^N \psi_i$ is the proportion of individuals with the trait. 
\end{proposition}

The inequality of (\ref{assumption2}) will be tighter if $BC$ comes from a study or studies of identical twins that were reared apart; see (\ref{oneone}). We prove Proposition \ref{probandwiseprop} in Appendix \ref{AppA}.

On a large or infinite population write $P(e=0,d=1)$, $P(e=1,d=1)$, $P(e=0,d=0)$, and $P(e=1,d=0)$ for the observed relative frequencies of a contingency table. Write $P(e=1)$ and $P(d=1)$ for the observed marginal relative frequencies of $e$ and $d$. Define the expected risk $r=(1-\pi)r_0+\pi r_1$. Write $m$ for a generic, variable distribution $m\in\mathcal P(C)$. We may partially identify the ATE of (\ref{ATE}) by solving the following constrained optimization problems. 
\begin{subequations}
\label{probs}
\begin{align}
\label{objective}
\textrm{ATE}_{\textrm{min/max}}  \  = \ 
\underset{m\in\mathcal P(C)}{\textrm{min/max}}\ & \int_C (r_{1}-r_{0}) dm
, \\ 
\label{cone}
\textrm{subject to } \ & 
\int_C (1-\pi)r_{0}dm \ = \ P(e=0,d=1), \\
\label{ctwo}
& \int_C \pi r_{1}dm \ = \ P(e=1,d=1), \\
\label{cthree}
& \int_C (1-\pi)(1-r_{0})dm \ = \ P(e=0,d=0), \\
\label{cfour} 
& \int_C \pi(1-r_{1})dm  \ = \ P(e=1,d=0),\\
\label{ione} 
& \int_C (\pi-P(e=1))^2dm \leq P(e=1)(BC_e-P(e=1)),  \textrm{~and} \\
\label{itwo} 
& \int_C (r-P(d=1))^2dm \leq P(d=1)(BC_d-P(d=1)).
\end{align}
\end{subequations}
Constraints (\ref{cone}-\ref{cfour}) require of a feasible distribution $m\in\mathcal P(C)$ that it must explain the observed relative frequencies. From those constraints we deduce the following mean identities $\bar{\pi}=P(e=1)$ and $\bar{r}=P(d=1)$. Those mean identities and the inequality of (\ref{assumption2}) justify the inequalities of (\ref{ione}) and (\ref{itwo}). 

The lower bound $\textrm{ATE}_{\min}$ represents the minimum ATE value that is consistent with the observed data, transported measures of concordance, and Assumption (\ref{assumption}). The upper bound $\textrm{ATE}_{\max}$ represents the maximum ATE value that is consistent with the observed data, transported measures of concordance, and Assumption (\ref{assumption}). 

We approximate both the minimization and maximization problems in \eqref{probs} by partitioning the cube $C$ and representing $m$ by a probability density function that is piecewise constant on each partition component. 
This results in a linear program, which we implement in python and solve using the \texttt{optimize.linprog} function in the \texttt{scipy} library.
\section{Example Applications}
\label{applications}
Here we apply our algorithm to example applications to demonstrate how the output of our algorithm can help researchers quantify uncertainty of causal interpretations from observed associations. We analyze associations between diabetes and stroke, smoking and chronic obstructive pulmonary disease (COPD), and marijuana use and subsequent use of harder drugs. We utilize transported measures of concordance from studies of identical twins. Transportability is an emerging area of interest in causal inference \cite{Mitra2022}, and there are subtleties when transporting concordances \cite[Fig. 2b]{Hagenbeek2023}. The following applications are not meant to definitively support any scientific conclusions, but rather to simply demonstrate our introduced methodology in contexts that are familiar to most readers. The stated results of the following example applications rely on the assumptions of Section \ref{methods}, including SUTVA.

\subsection{Does diabetes cause stroke?}
\begin{table}[h!]
\centering
\caption{A contingency table, from \cite{Steg}, showing an observed association (RR $=5.8$) between diabetes and stroke.}
\label{T1}
\begin{tabular}{rcc}
\toprule
 & \multicolumn{2}{c}{Diabetes}\\
\cmidrule{2-3}
Stroke&No&Yes\\
\hline
Yes&1,823&647\\
No&110,986&6,277\\
\hline
\end{tabular}
\end{table}

\noindent From the data of Table \ref{T1}, taken from \cite{Steg}, there is an observed association $RR=5.8$ between diabetes and stroke, but the exposure (diabetes) has not been randomly assigned. Medici et al. have studied diabetes in identical twins and reported a range of pairwise concordance rates \cite{Medici1999}. Based on the midpoint of their range we set $PC_e=0.45$. Equation \eqref{e:PCBC} of Appendix \ref{AppA} gives a formula for computing $BC$ from $PC$, and by that formula we obtain $BC_e=0.62$. We obtain $BC_d=0.18$ directly from \cite{Bak2002}. 
With those parameters and the data of Table \ref{T1} we solve \eqref{probs} resulting in
\[\textrm{ATE}_{\min}=-.01\leq \textrm{ATE}\leq 0.48=\textrm{ATE}_{\max}.\]
Because the interval $(-.01,.48)$ contains zero we have not demonstrated that the ATE is positive. It could be that diabetes does not cause stroke. Interventions that treat diabetes may not save lives that would otherwise be lost to stroke.

\subsection{Does smoking cause COPD?}
\label{smoking}

\begin{table}[ht!]
\centering
\caption{A contingency table, taken from \cite{Terz}, showing an observed association (Relative Risk, RR $=2.8$) between Cigarette Smoking and Chronic Obstructive Pulmonary Disease (COPD).}
\label{COPD}
\begin{tabular}{rcc}
\toprule
 & \multicolumn{2}{c}{Smoking}\\
\cmidrule{2-3}
COPD&No&Yes\\
\hline
Yes&318&1,631\\
No&4,679&7,538\\
\hline
\end{tabular}
\end{table}

\noindent From the data of Table \ref{COPD}, taken from \cite{Terz}, we can compute an observed association ($RR=2.8$) between cigarette smoking and Chronic Obstructive Pulmonary Disease (COPD), but the exposure (smoking) has not been randomly assigned. Among monozygotic twins, the probandwise concordance of smoking has been estimated to be about $BC_e=.67$ \cite{Kaprio1984}, and the probandwise concordance of COPD has been estimated to be about $BC_d=.20$ \cite{Ingebrigtsen2010}. With these parameters and the data of Table \ref{COPD} we solve \eqref{probs} resulting in
\[\textrm{ATE}_{\min}=.03\leq \textrm{ATE}\leq .21=\textrm{ATE}_{\max}.\]
The interval $(.03,.21)$ does not contain zero, implying that the ATE is positive. The interpretation is that individuals who choose to smoke increase their risk of COPD by at least $3\%$ on average. Interventions that coerce individuals to cease smoking are expected to reduce risk of COPD by at least $3\%$ and possibly by as much as $21\%$, on average.

\subsection{Does marijuana use cause harder drug use?}
\label{marijuana}
\begin{table}[ht!]
\centering
\caption{A contingency table, taken from \cite{PATH}, showing an observed association (RR $=11$) between marijuana use and hard drug use.}
\label{Drugs2}
\begin{tabular}{rcc}
\toprule
 & \multicolumn{2}{c}{Marijuana use}\\
\cmidrule{2-3}
Hard drug use&No&Yes\\
\hline
Yes&114&978\\
No&3,649&1,864\\
\hline
\end{tabular}
\end{table}

\noindent  The data of Table \ref{Drugs2} were taken from \cite{PATH}, without weighting. 
We defined the category of harder drugs to include cocaine, methamphetamine, speed, and heroin. From the data of Table \ref{Drugs2} we can compute $RR=11$, but the exposure (marijuana) has not been randomly assigned. We set $BC_e=0.50$ based on \cite{Kendler1998}, and we set $BC_d=.40$ based on \cite{Kendler1998b} and \cite{T96}. With these parameters and the data of Table \ref{Drugs2} we solve \eqref{probs} resulting in
\[\textrm{ATE}_{\min}=.11\leq \textrm{ATE}\leq .52=\textrm{ATE}_{\max}.\]
The interval $(.11,.52)$ does not contain zero, implying that the ATE is positive. The interpretation is that individuals who choose to use marijuana increase their risk of hard drug use by at least $11\%$ on average. Interventions that successfully convince individuals to avoid marijuana are expected to reduce the risk of hard drug use by at least $11\%$ and possibly by as much as $52\%$, on average.

\section{Discussion}
\label{discussion}
We have developed a method to partially identify the ATE from measured data and measures of twin concordance. Its inputs are observed relative frequencies $P(e=0,d=1)$, $P(e=1,d=1)$, $P(e=0,d=0)$, and $P(e=1,d=0)$, along with transported measures of probandwise concordances $BC_e$ and $BC_d$. Its outputs are lower and upper bounds for the ATE. Discordant twins imply randomness in the data generating process, resulting in tighter bounds on the ATE. Tighter bounds are important becaues more precise knowledge of the ATE can lead to more reliable intervention policy.

%

Our method is designed to extract useful information from simple and common contingency tables for the purpose of causal interpretation. If within a contingency table a cell count $k$ is small, say $k<100$, we can resample to assess sensitivity to our infinite population assumption. This approach is described in \cite[Section 3]{https://doi.org/10.48550/arxiv.2103.05692}. 
If there are measured covariates that information can be taken into account utilizing the approach described in \cite[Section 4]{https://doi.org/10.48550/arxiv.2103.05692}. The idea is to express the constraints on lines \eqref{cone}--\eqref{cfour} conditionally on the levels of the covariates. However, that approach is limited by the curse of dimensionality, and it can not be used to adjust for measured high-dimensional covariate data. We may apply our methodology conditionally on a sub population defined from measured covariates though. 
%

Covariate measurements should be made prior to exposure and at or prior to the time when propensity probabilities are defined. When utilizing twin studies to demonstrate randomness in the data generating process, we recommend defining propensity and prognosis probabilities at the moment of conception for each individual. That requirement can be relaxed in situations where randomness is demonstrated through other means. When propensity probabilities are defined long before the time of exposure, the possibility of multiple versions of the exposure arises, bringing SUTVA into question. The process that gives rise to marijuana use could also give rise to alcohol use, for example, and alcohol use could be an additional cause of hard drug use. This problem is not unique to our methodology. Multiple versions of the exposure may occur even when propensity probabilities are defined at times just before the exposure and even within experiments \cite{Hernan}. 

To check for a possible SUTVA violation we recommend transport of coefficients of determination from relevant studies. To continue the above example, Alexander et al. modeled drug use and reported pseudo $R^2$ values under $.30$ \cite{Alexander2017}. The type of $R^2$ reported was Tjur's $R^2$ \cite{Tjur2009}. In \cite[Section 3]{KnaEnt} we show that Tjur's $R^2$ measures the proportion of variance explained by varying individual propensity probabilities. Like the main argument in \cite{Cornfield}, since the study of Alexander et al. \cite{Alexander2017} was thorough and utilized numerous predictors, its relatively low $R^2$ values provide some evidence against chance confounding of the relationship between marijuana and hard drug use. By chance confounding we mean confounding by chance events that occur long after the times of definition for the propensity probabilities. Alexander et al. \cite{Alexander2017} utilized predictors defined and active closer to the times of exposure, and their lower $R^2$ values support a claim that the randomness \cite{Knaeble2023} that can be conclusively inferred from discordant twins (see Proposition \ref{probandwiseprop}), is likely still active right up until the exposure times and outcome times of our example application in Section \ref{marijuana}.

In \cite{Knaeble2023} we maximized a measure of randomness. In \cite{KnaEnt} we estimated the distribution of heterogeneous effects by maximizing entropy. Here we have utilized similar techniques to bound the ATE. Other causal estimands such as the 
ratio-based ATE, $\int_C r_1d\mu/\int_C r_0d\mu$, 
or a measure of causality present, 
$\int_C|r_1-r_0|d\mu$,  
could also be bounded by appropriately replacing the objective function in \eqref{objective}.

\clearpage 
\appendix
\section{Variance of propensity probabilities and measures of concordance in twin studies}
\label{AppA}
Consider a twin study and let \\
\indent $n$ be the number of twin pairs,\\
\indent $C$ be the count of pairs where both have the trait, \\ 
\indent $D$ be the count of twin pairs where one but not both have the trait, and \\ 
\indent $U=n-C-D$ be the number of pairs where neither have the trait. \\
Denote the probandwise concordance by $
BC= \frac{2C}{2C+D} 
$ and the pairwise concordance by 
$
PC= \frac{C}{C+D}.
$
We observe that 
\begin{equation}
\label{e:PCBC}
\frac{1-PC}{1+PC}=\frac{D/(C+D)}{(2C+D)/(C+D)}=\frac{D}{2C+D}=1-BC, 
\end{equation}
which can be used to obtain BC and PC from each other. 
Finally, denote the proportion of homogeneous pairs (either both or neither have the trait) by
$$
V = \frac{U+C}{n}.  
$$ 

\noindent {\bf Proof of Proposition~\ref{probandwiseprop}.}
Write $\sigma^2$ for the variance of the distribution of $\psi$ values:
\[\sigma^2:=\frac{1}{N} \sum_{i=1}^N(\psi_i-\bar{\psi})^2.\] Using assumption \eqref{assumption}, we compute 
\begin{align*}
V & \geq \frac{1}{n} \sum_{i=1}^n\psi_i^2 + (1-\psi_i)^2 \\ 
&= 2 \sigma^2 - 2 \bar{\psi} (1-\bar{\psi}) + 1. 
\end{align*}
On the other hand, we have that 
\begin{align*}
V = \frac{U+C}{n}  
& = 1-\frac{D}{n} \\
&= 1-2\left(\frac{2C+D/2}{n}\right)\left(\frac{D/2}{C+D/2}\right)\\
&= 1-2\left(\frac{2C+D}{2n}\right)\left(1 - \frac{2C}{2C+D}\right)\\ 
&= 1- 2\bar{\psi}\left(1-BC\right), 
\end{align*}
where we have used that the proportion of individuals with the trait is $\bar{\psi}= \frac{2C+D}{2n}$. 
Combining these two statements, we obtain 
$\sigma^2 \leq \bar{\psi}( BC - \bar{\psi})$,
as desired.
\hfill $\square$

\end{document}